**Semi-Nonparametric Estimation of Operational Risk Capital with Extreme Loss Events**

Heng Z. Chen[a], Stephen R. Cosslett[b]

a: Corresponding author, Northwestern University. Email: heng.chen@northwestern.edu and heng.zhang.chen@gmail.com.

b: The Ohio State University, Columbus, Ohio. Email: cosslett.1@osu.edu

**Keywords**
semi-nonparametric estimation; extreme value theory; value at risk; operational risks; Basel regulatory capital requirements

**Abstract**

Bank's operational risk capital modeling using the Basel II advanced measurement approach (AMA) often lead to a counter-intuitive capital estimate of value at risk at 99.9% due to extreme loss events. To address this issue, a flexible semi-nonparametric (SNP) model is introduced using the change of variables technique to enrich the family of distributions to handle extreme loss events. The SNP models are proved to have the same maximum domain of attraction (MDA) as the parametric kernels, and it follows that the SNP models are consistent with the extreme value theory - peaks over threshold method but with different shape and scale parameters from the kernels. By using the simulation dataset generated from a mixture of distributions with both light and heavy tails, the SNP models in the Fréchet and Gumbel MDAs are shown to fit the tail dataset satisfactorily through increasing the number of model parameters. The SNP model quantile estimates at 99.9% are not overly sensitive towards the body-tail threshold change, which is in sharp contrast to the parametric models. When applied to a bank's operational risk dataset with three Basel event types, the SNP model provides a significant improvement in the goodness of fit to the two event types with heavy tails, yielding an intuitive capital estimate that is in the same magnitude as the event type total loss. Since the third event type does not have a heavy tail, the parametric model yields an intuitive capital estimate, and the SNP model cannot provide additional improvement. This research suggests that the SNP model may enable banks to continue with AMA or its partial use to obtain an intuitive operational risk capital estimate when the simple non-model based Basic Indicator Approach (BIA) or Standardized Approach (TSA) are not suitable per Basel Committee Banking Supervision OPE10 (2019).

# 1. Introduction

Operational risk is defined as the risk of loss resulting from inadequate or failed internal processes, people, and systems or from external events (Basel Committee on Banking Supervision 2006). It has been identified as an increasingly important factor for bank performance and stability, and one which banks maintain substantial capital reserves against. In recent years, the public media has reported numerous very large operational loss events that exceed several billion dollars. This is because capital estimation based on loss history and measured as value at risk (VaR) can be critically influenced by extreme loss events, especially regulatory fines for bank's misconducts such as CPBP.[1] For examples, based on the data collected by the Basel Committee, Moscadelli (2004) found that by following the extreme value theory - peaks over threshold (EVT-POT) approach and partitioning the operational risk loss events into the body and tail, the Pareto tail shape parameter estimate often exceeds one, suggesting that the distribution has an infinite mean and/or variance with a counter-intuitive VaR estimate. Cope *et al.* (2009) conducted a sensitivity analysis and found that by removing the top three loss events from the modelling data sample, the quantile estimate at 99.9% reduces by a 65%, reflecting the significant impact of extreme loss events in the capital estimation.

To address the VaR sensitivity with respect to tail events, robust estimators have been introduced in the literature to address the concern about influence of outliers in the capital estimation. By "robustifying" maximum likelihood estimation, Choi *et al.* (2000) proposed to tilt the likelihood by choosing weights for the score function with a divergence index such as Kullback-Leibler with unequal priors. Colombo *et al.* (2015) applied the weighted maximum likelihood estimation approach in operational risk modeling. Horbenko *et al.* (2011) discussed the use of robust statistics such as the optimal mean square error estimator and the radius min-max estimator. A problem is that the radius of the optimal neighbourhood is never known, and finding a tuning parameter for the radius is a difficult task. This challenge is similar to the approach of weighted maximum likelihood or other robust statistics which requires a carefully constructed optimization problem with some form of subjective judgement in order to control the influence function or sensitivity to extreme events. As a consequence, assuming the extreme events as contaminated and/or the selection of tuning parameters in the model development can make the capital estimation vulnerable to gaming and manipulation.

---

[1] The operational loss events are categorized into seven event types by the Basel Committee, namely Internal Fraud (IF); External Fraud (EXTF); Employment Practices and Workplace Safety (EPWS); Clients, Products, and Business Practice (CPBP); Damage to Physical Assets (DPA); Business Disruption and Systems Failures (BDSF); and Execution, Delivery, and Process Management (EDPM).

The econometrics literature suggests that the semi-nonparametric (SNP) distribution is flexible and can be used to approximate any unknown data generating process, especially when the underlying distribution contains heavy tails for which there does not exist a suitable standard parametric distribution. Gallant and Nychka (1987) introduced the SNP methodology by combining a normal kernel with Hermite polynomials, and showed that the approximation errors can be made arbitrarily small by increasing the polynomial truncation point. Chen (2007) indicated that another attractive feature of the SNP methodology is its ease of implementation since the SNP distribution can often be characterized by a finite number of parameters, reduced to a parametric model, and thus estimated by maximum likelihood, generalized least squares, sieve minimum distance and other methods. However, to obtain the desired theoretical properties of the SNP estimator, it is necessary that the number of parameters increase slowly with the sample size. Ñíguez and Perote (2011) argued that the proposed SNP methodology for computing VaR is more accurate than the traditional Gaussian-assumption-based methods in RiskMetrics. Chen and Randall (1997) introduced a semi-nonparametric estimation using the change of variables technique in the context of binary choice models and demonstrated its asymptotic statistical properties. Since the SNP model nests any chosen parametric model as a special case with the mixed density functions and weights, its significance can be evaluated using the traditional model's goodness of fit test to detect any departure from the parametric distribution and/or existence of higher moments in the dataset. The paper demonstrated that the SNP model resulted in an improved fit to the dataset and a substantially different willingness to pay estimate for environmental amenity from that of the initial parametric model.

In this research, we extend the SNP estimation by change of variables to model tail events above the EVT-POT threshold in the capital estimation and demonstrate that the SNP model can yield an intuitive capital estimate without treating extreme loss events as contaminated data points. Fundamentally, we believe that extreme events in operational risk management are not actually contaminated data points. Rather, they have been accepted as legitimate loss events which have been carefully scrutinized by the business and modeling communities, even though they might not fit to a particular parametric distribution assumption. For that reason, our approach does not use robust estimators; instead, it is based on the view that excessive sensitivity to extreme loss events is mainly due to the loss distribution misspecification that might be caused by multiple data generating processes from the business lines with different characteristics that are either observable or unobservable from the modeler's perspective. As a result, this research suggests that the SNP model may enable banks to continue with AMA or its partial use to obtain an intuitive operational risk capital estimate that is sensitive to risk exposure, and to avoid the weaknesses caused by the use of Gross Income (GI) as a proxy indicator in the simple non-model based Basic Indicator Approach (BIA) or Standardized Approach (TSA).

After the introduction, Section 2 starts with a brief overview of the SNP estimation by change of variables. We prove that the SNP distribution under the polynomial series transformation has the same MDA as the kernel. Therefore, the SNP model is consistent with the EVT-POT approach, and it can yield an intuitive capital estimate with shape parameter that is a function of the kernel's shape parameter and polynomial series transformation. As a result, the SNP distribution enriches the family of distributions that can be used to handle heavy tails that the traditional parametric models cannot. In Section 3, in order to compare sensitivity of the quantile estimates at 99.9% with respect to the body-tail threshold between the parametric and the SNP models, three datasets are simulated from a mixture of data generating processes with both light and heavy tails. Based on the estimated sixty-three models with three different body-tail thresholds and distribution assumptions, it is observed that the SNP model quantile estimates are not overly sensitive towards the threshold change, which is in sharp contrast to the parametric model quantile estimates. In Section 4, an actual operational risk loss dataset with three Basel event types is selected to assess the SNP model suitability for the VaR capital estimation. It is found that for two of the three Basel event types with heavy or extremely heavy tails, the SNP model yields an intuitive and reasonable capital estimate that is in the same magnitude as the total loss. Since the third Basel event type has a regular tail with Pareto shape parameter less than one, the parametric model yields a reasonable capital estimate that is also in the same magnitude as the total loss. In this case, the SNP model could not provide a further improvement over the parametric model. Section 5 concludes the research.

**2. The SNP Estimation by Change of Variables**

In this section, we extend the SNP estimation by change of variables to model tail loss events in the EVT-POT approach, and prove that the SNP distribution has the same MDA as the chosen parametric kernel. Therefore, the SNP distribution enriches the family of distributions that may be used to approximate the true data generating process with both light and heavy tails through suitably increasing the number of parameters in the polynomial power series in the Fréchet, Gumbel, and Weibull MDAs. The SNP model's log-likelihood functions are also derived for the maximum likelihood estimation in the following Sections 3 and 4.

## 2.1. Regularity conditions and constrained power series

Let $f(x)$ be the density function of any continuous variable $x$, which can be obtained through a Jacobian transformation from variable $v$ with a known kernel density function $g(v)$. That is, if variable $v$ is a monotone increasing function of variable $x$, written as $v = h(x)$, then the density function $f(x)$ can be written as

$$f(x) = g(h(x)) \nabla_x h(x) \tag{1}$$

where the transformation is $\partial v/\partial x = \partial h(x)/\partial x \equiv \nabla_x h(x)$. Note that the transformation $h(\cdot)$ should satisfy the following regularity conditions of monotonicity and non-negativity to ensure that density function $f(x)$ is non-negative, and the corresponding cumulative distribution function is monotonically increasing between 0 and 1 for loss amount $x > 0$:

$\nabla_x h(x) \geq 0$

$h(x) \geq 0.$

As an example, the following power series will guarantee the above conditions with the gradient function as follows:

$$\nabla_x h(x, \theta_0, \theta_1, \cdots, \theta_K) = \left(\sum_{k=0}^{K} \theta_k x^k\right)^2 \equiv \sum_{i=1}^{m} i\gamma_i x^{i-1} \geq 0 \tag{2}$$

where $m = 2K + 1$. Thus, the SNP transformation $h(x)$ can be approximated by

$$h(x, \theta_0, \theta_1, \cdots, \theta_K) = \int_0^x \left(\sum_{k=0}^{K} \theta_k \eta^k\right)^2 d\eta \equiv \sum_{i=1}^{m} \gamma_i x^i \tag{3}$$

where $K$, the truncation point, (or equivalently $m$, the order of the power series) is determined by assessing if the additional parameter leads to an improved model fit.

For examples, if the power series is truncated at $K = 1$, the relationship between $\gamma$ and $\theta$ for the SNP distribution with one additional parameter is as follows:

$\gamma_1 = \theta_0^2$, $\gamma_2 = \theta_0 \theta_1$, and $\gamma_3 = \frac{1}{3}\theta_1^2$

- 5

If the power series is truncated at $K = 2$, the relationship between $\gamma$ and $\theta$ for the SNP distribution with two additional parameters is as follows:

$$\gamma_1 = \theta_0^2, \gamma_2 = \theta_0\theta_1, \gamma_3 = \frac{1}{3}(\theta_1^2 + 2\theta_0\theta_2), \gamma_4 = \frac{1}{2}\theta_1\theta_2, \text{ and } \gamma_5 = \frac{1}{5}\theta_2^2$$

If the power series is truncated at $K = 3$, the relationship between $\gamma$ and $\theta$ for the SNP distribution with three additional parameters is given by

$$\gamma_1 = \theta_0^2, \gamma_2 = \theta_0\theta_1, \gamma_3 = \frac{1}{3}(\theta_1^2 + 2\theta_0\theta_2), \gamma_4 = \frac{1}{2}(\theta_1\theta_2 + \theta_0\theta_3),$$

$$\gamma_5 = \frac{1}{5}(\theta_2^2 + \theta_1\theta_3), \gamma_6 = \frac{1}{3}\theta_2\theta_3, \text{ and } \gamma_7 = \frac{1}{7}\theta_3^2 \qquad (4)$$

The relationship between $\gamma$ and $\theta$ for $K \geq 4$ can be similarly derived.

*2.2. The SNP distribution and maximum domain of attractions*

To ensure that the SNP model by change of variables is consistent with the extreme value theory, this section proves that the SNP distribution $F(x) = G(h(x))$ with the power series under monotonicity belongs to the same MDA as the kernel distribution function $G$ for each of Fréchet, Gumbel, and Weibull MDAs. In general, to determine the MDA of SNP distribution $F$ that is a function of the power series under monotonicity, we can apply the method used in Example 3.3.31 in Embrechts *et al.* (1997) as follows.

Let $h(x)$ be a polynomial of degree $m \geq 1$ in (3),

$$h(x) = \sum_{i=0}^{m} \gamma_i x^i, \quad \gamma_m > 0,$$

subject to $h(x) > 0$ and $dh(x)/dx > 0$ for $x > x_0$. The tail index is not affected by a scale factor, so without loss we can set $\gamma_m = 1$. For large $x$ we then have

$$h(x) = x^m(1 + O(x^{-1})).$$

Since $h(x)$ is monotonic, there is an inverse function $r(y) \equiv h^{-1}(y)$, which has the following properties:

$$r(y) = y^{\frac{1}{m}} + O(1)$$

$$r'(y) = \frac{1}{h'(r(y))} = O\left(y^{-1+\frac{1}{m}}\right)$$

$$r''(y) = -\frac{h''(r(y))}{h'(r(y))^3} = O(y^{-2+\frac{1}{m}})$$

- 6

Let $F(x) = G(h(x))$, where $G$ has the MDA of distribution $H$. Let $M_n$ be the maximum of $n$ i.i.d. random variables $Y \sim G$ and let $\widetilde{M}_n$ be the corresponding maximum for random variables $X \sim F$. Suppose $c_n$ and $d_n$ are such that

$$P\{M_n(G) \leq c_n z + d_n\} \to H(z).$$

Since $h(x)$ is monotonic,

$$\widetilde{M}_n = r(M_n)$$

and therefore

$$P\{\widetilde{M}_n \leq r(c_n z + d_n)\} \to H(z).$$

We can then apply the method used in Example 3.3.31 in Embrechts *et al*. (1997) to determine the MDA of $F$.

***Result* 1**. If $G$ belongs to the Gumbel MDA, then $F(\cdot) = G(h(\cdot))$ also belongs to the Gumbel MDA.

By the mean value theorem,

$$r(c_n z + d_n) = r(d_n) + c_n z r'(d_n) + \lambda (c_n z)^2 r''(d_n)$$

for some $\lambda \in [0,1]$, and therefore

$$P\left\{\frac{\widetilde{M}_n - r(d_n)}{c_n r'(d_n)} \leq z + O\left(\frac{c_n}{d_n}\right)\right\} \to \Lambda(z).$$

Since the Gumbel MDA consists of von Mises functions and distribution functions that are tail-equivalent to von Mises functions, we need only consider the case where $G$ is a von Mises function. Denoting the corresponding auxiliary function by $a(\cdot)$, suitable scaling parameters are $d_n = G^{-1}(1 - n^{-1})$ and $c_n = a(d_n)$. Let $x^* \leq \infty$ be the upper bound of the support of $G$. There are two cases to consider. (a) If $x^* = \infty$, then $d_n \to \infty$ and $a(x)/x \to 0$ as $x \to \infty$, from which it follows that $c_n/d_n \to 0$. (b) If $x^* < \infty$, then $d_n \to x^*$ and $a(x)/(x^* - x) \to 0$ as $x \uparrow x^*$. But for large enough $n$, $|x^* - d_n| \leq |d_n|$ and therefore $c_n/d_n \to 0$ in this case also. Since $c_n/d_n \to 0$ and $\Lambda$ is continuous, it follows that

$$P\{\widetilde{M}_n \leq \tilde{c}_n z + \tilde{d}_n\} \to \Lambda(z)$$

with $\tilde{c}_n = c_n r'(d_n)$ and $\tilde{d}_n = r(d_n)$, and thus $F$ also belongs to the Gumbel MDA. □

***Result* 2.** If $G$ belongs to the Fréchet MDA $\Phi_\alpha$, then $F(\cdot) = G(h(\cdot))$ also belongs to the Fréchet MDA $\Phi_{m\alpha}$.

In this case, $c_n \to \infty$ and $d_n \to 0$. Then
$$P\{\widetilde{M}_n \leq r(c_n z)\} \to \Phi_\alpha(z)$$
implies
$$P\left\{\widetilde{M}_n \leq (c_n z)^{\frac{1}{m}} + O(1)\right\} \to \Phi_\alpha(z)$$
and therefore
$$P\left\{c_n^{-1/m} \widetilde{M}_n \leq z^{\frac{1}{m}} + O(c_n^{-1/m})\right\} \to \Phi_\alpha(z).$$
Since $c_n \to \infty$ and $\Phi_\alpha$ is continuous, it follows that
$$P\{\widetilde{M}_n \leq \tilde{c}_n z\} \to \Phi_\alpha(z^m) = \Phi_{m\alpha}(z)$$
with $\tilde{c}_n = c_n^{1/m}$, and thus $F$ belongs to the Fréchet MDA $\Phi_{m\alpha}$. □

**Result 3**. If $G$ belongs to the Weibull MDA $\Psi_\alpha$, then $F(\cdot) = G(h(\cdot))$ also belongs to the Weibull MDA $\Psi_\alpha$.

In this case, $c_n \to \infty$ and $d_n = x^*$. By the mean value theorem,
$$r(c_n z + x^*) = r(x^*) + c_n z r'(x^*) + \lambda (c_n z)^2 r''(x^*)$$
for some $\lambda \in [0,1]$, and thus
$$P\{\widetilde{M}_n \leq r(c_n z + x^*)\} \to \Psi_\alpha(z)$$
implies
$$P\left\{\frac{\widetilde{M}_n - r(x^*)}{c_n r'(x^*)} \leq z + O(c_n)\right\} \to \Psi_\alpha(z).$$
Since $c_n \to 0$ and $\Psi_\alpha$ is continuous, it follows that
$$P\{\widetilde{M}_n \leq \tilde{c}_n z + \tilde{x}^*\} \to \Psi_\alpha(z)$$
with $\tilde{c}_n = c_n r'(x^*)$ and $\tilde{x}^* = r(x^*)$, and thus $F$ belongs to the Weibull MDA $\Psi_\alpha$. □

*2.3. SNP model log-likelihood functions*

Fisher-Tippett Theorem stipulates that the asymptotic distribution of the maxima of independent identically distributed random variables will converge to one of the three maximum domain of attractions (MDA), namely the Fréchet, Weibull, and Gumbel MDA. Embrechts *et al.* (1997) and Alves and Neves (2017) list some of the distributions that fall into each MDA. Since the capital analysis is bank's proprietary information, it is difficult to obtain a comprehensive list of distributions used by industry practitioners. As a result, we selected a few simple and frequently used parametric distributions that fall into each of the three MDAs with both light and heavy tails. This section derives the SNP model log-likelihood functions based on the selected parametric kernel distributions.

Specifically, we selected Generalized Pareto (GPD) and Log-Logistic (LogLGT), Lognormal (LGN), Weibull (WBL), and Exponential (EXP) as kernels from the Fréchet, Gumbel, and Weibull MDAs. The cumulative distribution function for the GPD, LogLGT, and WBL models is given by $G(v, b, c) = 1 - (1 + c\frac{v}{b})^{-\frac{1}{c}}$, $G(v, b, c) = (v/b)^c (1 + (v/b)^c)^{-1}$, and $G(v, c, b) = 1 - \exp(-bv^c)$, respectively, where $v > 0$. The shape and scale parameters are given by $b > 0$ and $c > 0$, respectively. EXP is a special case of WBL by restricting $c = 1$. Therefore, the SNP models based on GPD, LogLGT, LGN, and WBL kernels will have the following $f(x)$ as probability density functions

$$\begin{aligned}
\text{SNPGPD:} \quad & f(x) = \nabla_x h(x)(1 + ch(x))^{-(1+1/c)}, \\
\text{SNPLGT:} \quad & f(x) = c\nabla h(x) h^{c-1}(x)(1 + h^c(x))^{-2}, \\
\text{SNPLGN:} \quad & f(x) = \frac{1}{c\sqrt{2\pi}h(x)} \exp\left(-\frac{(\log(h(x))-\mu)^2}{2c^2}\right) \nabla_x h(x), \\
\text{SNPWBL:} \quad & f(x) = c\nabla_x h(x) h^{c-1}(x) \exp(-h^c(x)).
\end{aligned} \tag{5}$$

respectively, where $c$ is the identifiable shape parameter for the kernel GPD, LogLGT, and WBL, while the scale parameter $b$ is absorbed by $\theta_0, \theta_1, \theta_2, \ldots$ in the transformation $h(x, \theta)$. For the SNPLGN model, $c$ is identifiable but $\mu$ is absorbed by the transformation $h(x, \theta)$. The density function $f(x) = 0$ if not tails.

By substituting $\nabla_x h(x)$ in (2) and $h(x)$ in (3), and suitably increasing the order of power series $m$, the log-likelihood functions for the SNPGPD, SNPLGT, SNPLGN, and SNPWBL models are given as follows

$$\log L_{snpgpd}(c, \theta | x) = \log\left\{\sum_{j=1}^{m} j\gamma_j x^{j-1}\right\} - \left(1 + \frac{1}{c}\right) \log\left\{1 + c \sum_{i=1}^{m} \gamma_i x^i\right\},$$

$$\log L_{snplgt}(c, \theta | x) =$$
$$\log(c) + \log\left\{\sum_{j=1}^{m} j\gamma_j x^{j-1}\right\} + (c-1)\log\left\{\sum_{i=1}^{m} \gamma_i x^i\right\} - 2\log\left\{1 + \left(\sum_{i=1}^{m} \gamma_i x^i\right)^c\right\},$$

$$\log L_{snplgn}(c, \theta | x) =$$
$$-\frac{1}{2}\log(2\pi c^2) - \log\left\{\sum_{i=1}^{m} \gamma_i x^i\right\} - \frac{1}{2c^2}\left\{\log\left(\sum_{i=1}^{m} \gamma_i x^i\right)\right\}^2 + \log\left\{\sum_{j=1}^{m} j\gamma_j x^{j-1}\right\},$$

$$\log L_{snpwbl}(c, \theta|x) =$$
$$\log(c) + \log\{\sum_{j=1}^{m} j\gamma_j x^{j-1}\} + (c-1)\log\{\sum_{i=1}^{m} \gamma_i x^i\} - \{\sum_{i=1}^{m} \gamma_i x^i\}^c \qquad (6)$$

respectively, where $\gamma = (\gamma_1, \ldots, \gamma_m)$ are related to the underlying parameters $\theta = (\theta_0, \ldots, \theta_K)$ as in (4). Since the SNP model nests the selected parametric model as a special case, traditional model specification tests such as the nested LR test and Student t-test test can be carried out. Furthermore, the Q-Q plot will be evaluated to ensure that the model does not over-predict or under-predict the observed, especially for tail events. The model's tail distributions $1 - F(x)$ will be compared given its critical influence on the capital estimation. Note that the SNP distribution is restricted to the positive domain for the operational risk loss events, unlike the *g*-and-*h* distribution which can be both positive and negative (Dutta and Perry, 2007).

*2.4. Aggregate loss distribution for VaR capital*

For the SNP model, loss severity distribution at quantile $1 - a$ is given by
$$x_a = h^{-1}(\tilde{a}(c), \theta),$$
where $\tilde{a}(c)$ is defined as $(a^{-c} - 1)/c$, $(1/a - 1)^{1/c}$, and $(-\log(a))^{1/c}$ for the SNPGPD, SNPLGT and SNPWBL models, respectively. The SNPLGN distribution quantile can also be numerically estimated even though its cumulative distribution function does not have an analytical expression. Following industry practice, the capital VaR at 99.9% for a particular Basel event type can be estimated by Monte Carlo simulation through convolution of the estimated SNP distribution for loss severity and a count distribution such as Poisson for loss frequency.[2]

**3. An Evaluation of SNP Models on Simulated Dataset**

In this section, we first generated a simulated dataset from a mixture of multiple data generating processes using Pareto, Log-Logistic, and Weibull distributions to resemble actual operational risk loss events with both light and heavy tails from business units with different characteristics. To evaluate the SNP model sensitivity of the quantile estimates at 99.9% with respect to the body-tail cutoff thresholds in the EVT-POT approach, three tail exceedance datasets are created by setting the threshold at 50, 30, and 10 on the simulated dataset, respectively. We then estimated the SNP models with five kernel distributions on the three tail datasets that contain 2.4%, 3.9% and 8.6% of the total sample size. For benchmarking, six popular parametric models with the number of parameters ranging from one to four are estimated. The model comparison criteria include the model's goodness of fit to the dataset, Q-Q plots, tail distributions, and quantile estimates at 99.9% that can critically impact the capital estimation in the VaR approach. The research suggests that the SNP models based on Pareto, Log-Logistic and Lognormal kernels in the Fréchet and Gumbel MDAs can provide a similar good fit to heavy-tailed datasets by gradually increasing the number of parameters in the polynomial power series. The SNP model quantile estimates at 99.9% are quite insensitive towards the change of the body-tail cutoff thresholds. On the other hand, the parametric model quantile estimates are highly sensitive to the cutoff thresholds. Furthermore, the SNP models based on Weibull and Exponential kernels in the Weibull MDA for maxima yield a poor fit to heavy-tailed simulation datasets due to its limitation of finite support of right end point. Increasing the number of SNP model parameters could not mitigate the Weibull MDA limitation in this case.

---

[2] Correlations among Basel event types can also be taken into account by copula models in the simulation of overall aggregate loss distribution with diversification benefits.

*3.1. Simulated dataset*

Under certain regularity conditions, the Pickands-Balkema-de Haan Theorem describes the limit distribution of scaled excess losses over a sufficiently high threshold, suggesting that the limit distribution will be a Pareto distribution. Thus, applying EVT to POT dataset entails choosing a sufficiently high threshold to divide the operational risk events into a body and a tail, thus fitting GPD to the scaled excess losses in the tail. This approach has been widely accepted in insurance and financial risk modeling under the assumption that data generating process is independent and identically and distributed (i.i.d.). When the number of tail observations is limited, EVT is often employed to justify the extrapolation and capital estimation at high quantiles. See for example Dutta and Perry (2007), Franzetti (2011), and others. However, the POT data may be generated from a mixture of multiple data generating processes with unobservable heteroscedasticity, and empirically it could be difficult to set the body-tail threshold as well. As indicated by Embrechts *et al.* (1997), if the threshold is set too low, it will introduce significant bias for the EVT-POT approach. If the threshold is set too high, there will not be enough observations in the tail, leading to elevated variance.

To reflect the actual loss events with both light and heavy tails that might come from business units with different loss characteristics, a simulated dataset with 3000 observations is generated from Weibull, Pareto, and Log-Logistic distributions. To investigate the SNP model's suitability for the capital estimation that critically depends on tail behaviors, three thresholds at 50, 30, and 10 are used to split the simulated dataset into body and tail events. Table 3.1 below presents the summary statistics of three modeling datasets of so called peaks over threshold or exceedance after subtracting the threshold from the tails.

| Table 3.1 Data Summary | Distribution | Count | Minimum | Mean | Maximum |
|---|---|---|---|---|---|
| Cut at 50 (2.4% Sample) | Weibull | 49 | 4.96 | 127.06 | 928.28 |
|  | Pareto | 13 | 20.56 | 957.43 | 7805.99 |
|  | Log-Logistic | 10 | 3.10 | 2033.75 | 16609.29 |
| Cut at 30 (3.9% Sample) | Weibull | 82 | 0.87 | 91.53 | 948.28 |
|  | Pareto | 21 | 0.89 | 608.02 | 7825.99 |
|  | Log-Logistic | 15 | 1.74 | 1371.03 | 16629.29 |
| Cut at 10 (8.6% Sample) | Weibull | 179 | 0.45 | 55.30 | 968.28 |
|  | Pareto | 47 | 0.09 | 284.15 | 7845.99 |
|  | Log-Logistic | 32 | 0.23 | 654.89 | 16649.29 |

For example, when the threshold is set at 50, Table 3.1 indicates that there are in total 72 tail events that exceed the threshold with 49, 13, and 10 events from Weibull, Pareto, and Log-Logistic distributions, respectively. Figure 3.1 below illustrates that both of the Log-Logistic and Pareto dataset are heavy-tailed with the largest event that is more than five times of the second largest event. On the other hand, although 49 of 72 events in the dataset is from the Weibull distribution, its largest event is much smaller than that from the Pareto and Log-Logistic distributions, indicating that the Weibull distribution has a light or short tail.

Figure 3.1: Top tail events from the Weibull, Pareto, and Log-Logistic distributions with threshold 50

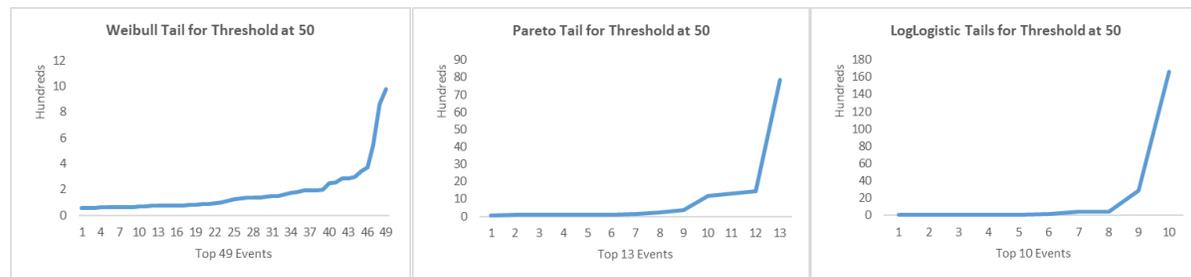

## 3.2 Model's goodness of fit and quantile estimates at 99.9%

To enable the model comparison on the dataset with both light and heavy tails, six parametric models with one to four parameters are estimated as the benchmark, namely Exponential, Pareto,

Log-Logistic, Lognormal, Weibull, and Generalized Beta of Type 2 (GB2). Furthermore, fifteen SNP models are estimated with the likelihood functions in (6) based on five parametric kernels with additional two, three and four parameters (2p, 3p, 4p) in the polynomial series in (3). Table 3.2 in the following presents the model performance in log-likelihood values and 99.9% quantile estimates for the sixty-three models on the three datasets with thresholds at 50, 30, and 10.

| Table 3.2 Model Performance[3] | Log-Likelihood | | | Quantile at 99.9% | | |
|---|---|---|---|---|---|---|
| Data Thresholds | 50 | 30 | 10 | 50 | 30 | 10 |
| GPD | 44.30 | 146.47 | 532.96 | 90,206 | 95,713 | 27,301 |
| LogLGT | 44.69 | 146.32 | 532.81 | 38,945 | 54,351 | 19,733 |
| LGN | 43.90 | 146.12 | 530.51 | 13,429 | 13,032 | 5,148 |
| WBL | 29.55 | 128.87 | 491.23 | 8,212 | 5,493 | 2,374 |
| EXP | -27.90 | 7.20 | 197.12 | 3,743 | 2,391 | 1,184 |
| GB2 | 47.83 | 147.17 | 533.15 | 205,644 | 44,804 | 20,877 |
| SNPGPD2p | 46.61 | 147.90 | 534.53 | 21,085 | 22,470 | 20,882 |
| SNPLGT2p | 47.46 | 147.88 | 534.51 | 21,832 | 23,283 | 21,205 |
| SNPLGN2p | 46.85 | 148.34 | 533.70 | 18,770 | 18,813 | 15,680 |
| SNPWBL2p | 41.67 | 142.55 | 514.41 | 17,042 | 16,392 | 2,648 |
| SNPEXP2p | 29.91 | 101.98 | 393.74 | 7,659 | 1,456 | 623 |
| SNPGPD3p | 47.33 | 149.36 | 535.88 | 18,993 | 18,636 | 17,370 |
| SNPLGT3p | 48.10 | 149.27 | 535.81 | 18,973 | 18,505 | 17,374 |
| SNPLGN3p | 49.13 | 150.32 | 535.93 | 17,735 | 17,801 | 16,940 |
| SNPWBL3p | 42.16 | 143.19 | 516.71 | 30,370 | 28,732 | 9,841 |
| SNPEXP3p | 33.62 | 110.84 | 425.54 | 14,378 | 1,628 | 578 |
| SNPGPD4p | 47.34 | 149.36 | 535.88 | 18,636 | 18,497 | 17,215 |
| SNPLGT4p | 48.18 | 149.28 | 535.83 | 18,325 | 18,241 | 17,152 |
| SNPLGN4p | 49.61 | 150.35 | 536.06 | 17,727 | 17,649 | 16,850 |
| SNPWBL4p | 42.82 | 143.87 | 518.58 | 17,438 | 17,014 | 10,087 |
| SNPEXP4p | 35.91 | 115.98 | 432.30 | 16,970 | 2,015 | 568 |

From Table 3.2, we observe that

1) Tail behaviors of the six parametric models are different significantly from each other and across the three datasets. For example, the GB2 model has the largest quantile estimate at threshold 50, but the GPD model has the largest quantile estimate at thresholds 30 and 10. The EXP model clearly has the lightest tail with the smallest quantile estimate across the three datasets. On the model performance, the GB2 models with four parameters outperform the models with one or two parameters (EXP, GPD, LogLGT, LGN, and WBL) in the log-likelihood value for the three datasets. The two parameters GPD and LogLGT models are doing reasonably well for some of the datasets, and the LGN model is clearly outperformed by the GPD and LogLGT models. This is simply because heavy tails in the simulated dataset are from the Pareto and Log-Logistic distributions. Although there is a large percentage of the POT tails from the Weibull distribution, the WBL model for maxima is not doing well due to its finite right end support that cannot accommodate heavy tails originated from the Pareto and Log-Logistic distributions. The one parameter EXP model performs the worst, which is included here just for the completeness and used as the SNPEXP model kernel.

---

[3] The model name suffix 2p, 3p, 4p means the SNP polynomial power series with additional two, three, and four parameters, respectively. For example, SNPLGN3p has five parameters with two from the LGN kernel and three from the SNP power series. The appendix provides the detailed sixty-three model parameter estimates and their log-likelihood values as well as the simulated dataset summary statistics.

2)      All of the SNP models outperform their parametric counter parts (EXP, GPD, LogLGT, LGN, and WBL), indicating that the power series (3) are quite significant for the approximation of Jacobian transformation. For example, the SNPGPD2p, SNPLGT2p, and SNPLGN2p models also outperform the GB2 models for the datasets with thresholds 30 and 10. But for threshold 50, the reverse is true. This suggests that there is a need to further improve the SNP models by increasing the power series truncation point.

3)      It turns out that the SNPLGN3p model provides a reasonable good fit, closely followed by the SNPGPD3p and SNPLGT3p models. The SNP models with four additional parameters have the performance similar to the SNP models with three additional parameters. The SNP models based on Weibull and Exponential kernels are not suitable for heavy-tailed dataset, limited by the Weibull MDA's finite support of right end points. Their quantile estimates are also quite volatile across the three datasets.

*3.3 Q-Q plots and tail behaviors*

Given the critical influence of tail behaviors on the capital estimation, the Q-Q plots for extreme tail events should be carefully observed in order to ensure that the model prediction is as close to the observed as possible along the 45-degree line. For examples,

1)      The Q-Q Plots in Figure 3.2a are for the six parametric models. It suggests that the two parameters GPD model performance is comparable to or better than the four parameters GB2 model on predicting extreme events. This is likely because the simulation dataset contain extreme events generated from the Pareto distribution with heavy tails. The remaining four parametric models are not performing as well as the GPD and GB2 models. That is, the LGN model is not doing as well as the GPD model across the three datasets. The WBL and EXP models clearly fall behind due to the Weibull MDA's limitation of finite right end support for the maxima.

Figure 3.2a Q-Q plots for the six parametric models at thresholds 50, 30, and 10

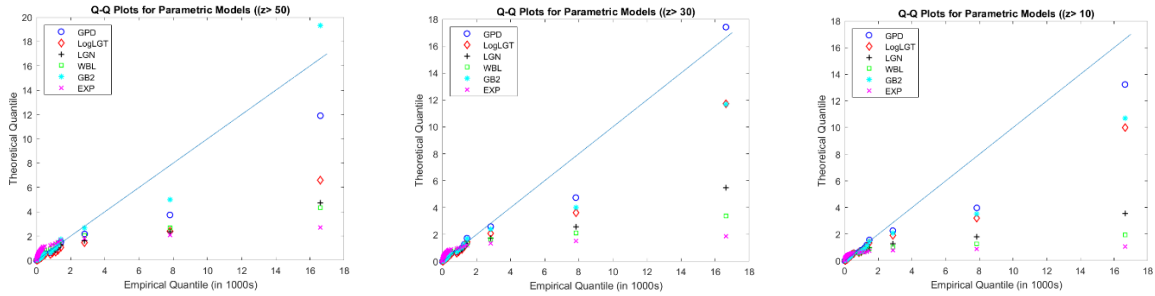

2)      Although we have observed that the SNPGPD, SNPLGT, and SNPLGN models with two additional parameters outperform their parametric counter parts, and are comparable to the four parameters GB2 model in term of the log-likelihood values, the Q-Q Plots in Figure 3.2b clearly indicate that there is room for improvement for the SNP models in predicting extreme tail events, especially at thresholds 30 and 10.

Figure 3.2b Q-Q plots for the SNP models with 2 additional parameters

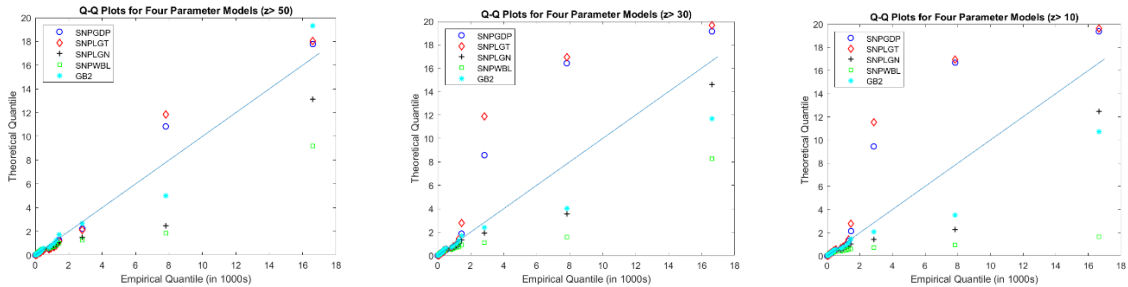

3)      Indeed, the Q-Q plots in Figure 3.3a suggest that all of the SNPGPD, SNPLGT, and SNPLGN models with three and four additional parameters in the Fréchet and Gumbel MDAs can provide a good fit to extreme tail events with the points close to the 45-degree line. In particular, the performance is comparable across the three datasets for the SNPLGN3p and SNPLGN4p models, both of which provide a slightly better fit than the other SNP models with the same number of parameters in the log-likelihood values and Q-Q plots for extreme tail events.

Figure 3.3a: Q-Q plots for the GPD and the SNPGPD models with 2, 3, and 4 additional parameters.

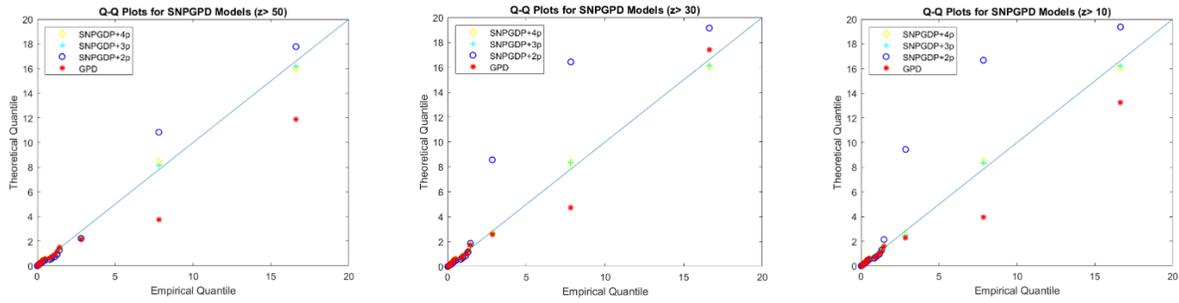

Figure 3.3b: Tail distributions for the GPD and SNPGPD models with 2, 3, and 4 additional parameters.

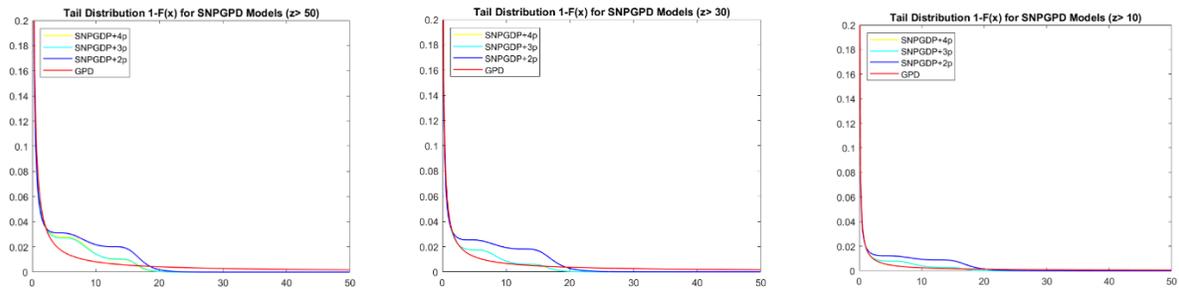

Figure 3.4a: Q-Q plots for the LogLGT and the SNPLGT models with 2, 3, and 4 additional parameters.

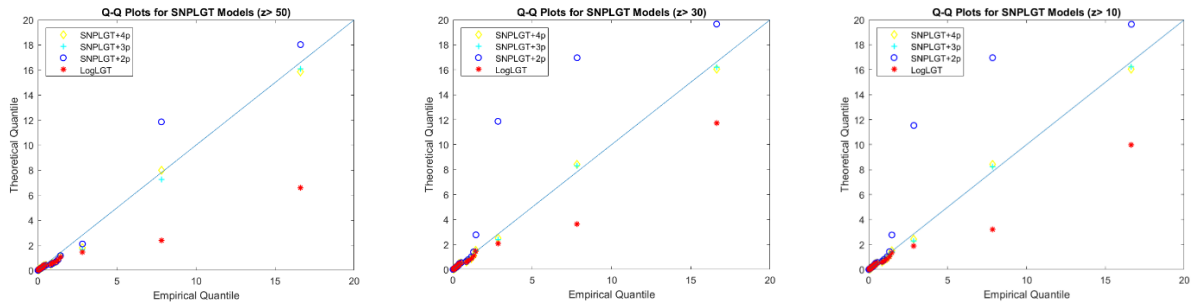

Figure 3.4b: Tail distributions for the LogLGT and the SNPLGT models with 2, 3, and 4 additional parameters.

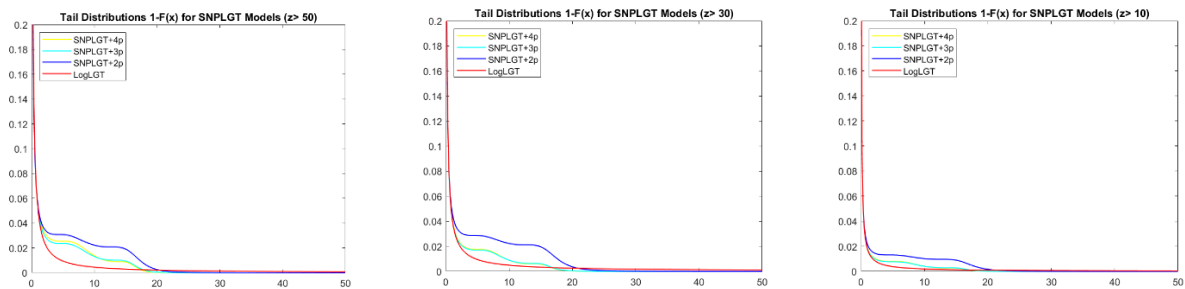

Figure 3.5a: Q-Q plots for the LGN and the SNPLGN models with 2, 3, and 4 additional parameters.

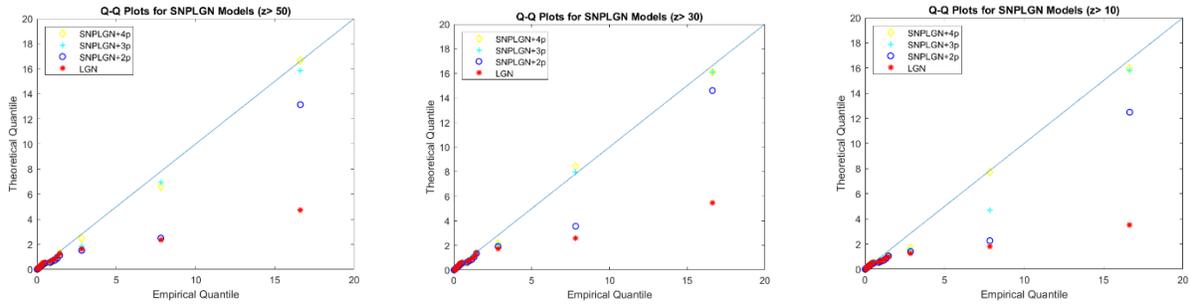

Figure 3.5b: Tail distributions for the LGN and the SNPLGN models with 2, 3, and 4 additional parameters.

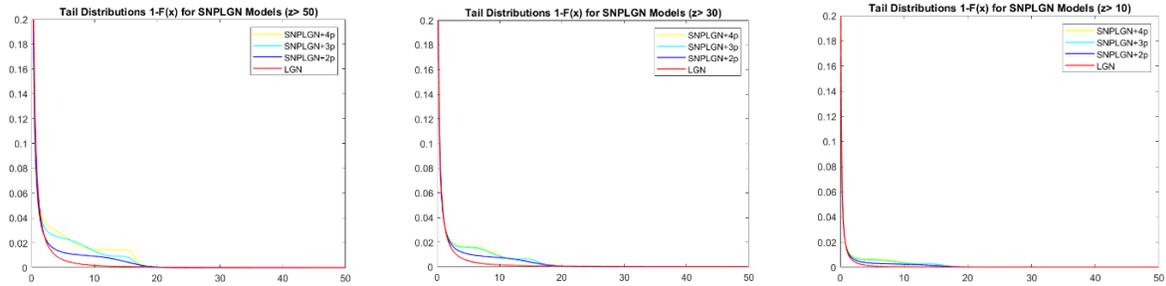

4) Although the simulation datasets in Table 3.1 include a large percentage of observations from the Weibull distribution, the SNPWBL and SNPEXP models with the increasing truncation point still under predict the extreme events due to its limitation of finite right end support for the Weibull MDA for maxima, as shown in the Q-Q plots in Figures 3.6 and 3.7.

Figure 3.6: Q-Q plots for the WBL and the SNPWBL models with 2, 3, and 4 additional parameters.

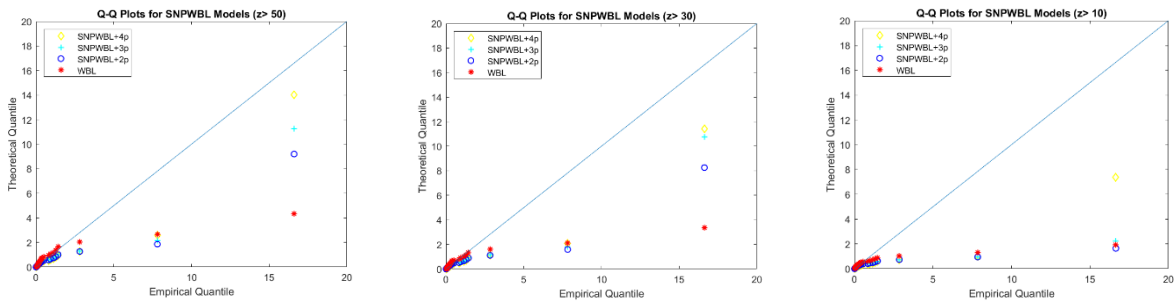

Figure 3.7: Q-Q plots for the EXP and the SNPEXP models with 2, 3, and 4 additional parameters.

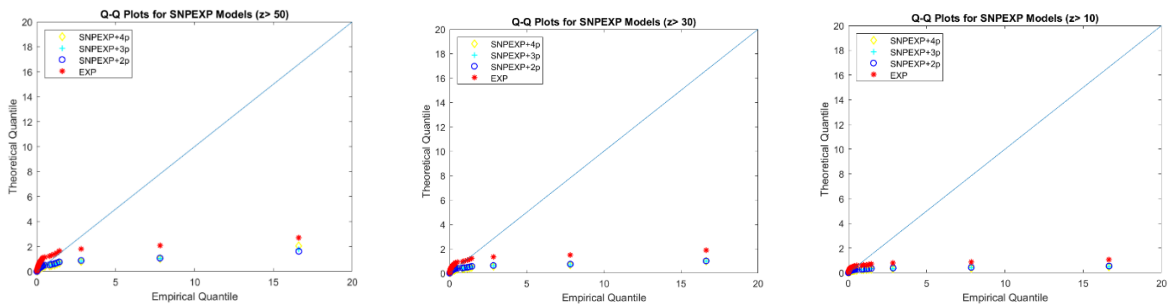

## 4. An SNP Model Application to Operational Risk Capital Modeling

This section illustrates the SNP model capital estimation using an actual operational risk loss dataset from a major international bank by the EVT-POT approach. The Basel framework did not specify the approach or distributional assumptions used to generate the operational risk measure for regulatory capital purposes, but a bank must be able to demonstrate that its approach captures the severely heavy-tailed loss events using the advanced modeling approach (AMA).[4] Neslova *et al.* (2006) raised concerns about naïve application of the EVT-POT approach to operational risk capital modelling without a careful understanding of the loss dataset. For example, the assumption of independent and identically distributed loss events may be difficult to maintain in practice. There might exist different data generating processes and/or different data reporting thresholds in the operational risk management. The true distribution might be heteroscedastic and unknown to modellers. The authors suggest that mixed true data generating processes can turn out to be difficult to detect if one does not look for them. Abdymomunov and Curti (2019) proposed to rescale the bank loss by total assets to arrive at a more stable capital estimate through combining peer banks' data to address heterogeneous losses caused by observable characteristics. In practice, however, there could exist unobservable heterogeneous characteristics across business units, at different time points, and for idiosyncratic reasons. Inspired by the insight and the flexible SNP model functional form with mixed density functions, we propose to investigate whether the SNP model can lead to a sensible capital estimation in the presence of extreme events due to unobservable heterogeneity in the operational risk capital modeling.

*4.1. Modelling data*

Bank operational risks are known for heavy-tailed loss distributions across Basel categories. The cause of extreme loss events may be different and not fully observable due to, for example, attorney client privileges. To investigate the SNP methodology suitability for the operational risk capital estimation in the presence of extreme loss events, a modeling dataset is selected from an international bank which consists of three Basel event types CPBP, EXTF and EDPM with minimum reporting loss amount of $10,000 USD between 2008Q1 and 2017Q1. The selected three event types account for more than 96% of the total loss, and the remaining four event types only account for less than 4% of the total loss during the period. Table 4.1 below provides the summary statistics for the selected event types, each of which is scaled by a constant to illustrate the SNP methodology for data conditioning and data confidentiality.

---

[4] Basel Committee Banking Supervision OPE10 (2019) will not mandate advanced measurement approach (AMA) starting 2020, provided banks satisfy certain conditions. However, banks may continue AMA, including partial use in business lines and event types.

| Table 4.1 Basel Event Types | Counts | Mean | Std Dev | Skewness | Maximum |
|---|---|---|---|---|---|
| External Fraud (EXTF) | 423 | 0.1 | 1.51 | 20.46 | 30.93 |
| Clients, Products and Business Practices (CPBP) | 324 | 0.1 | 1.07 | 14.13 | 17.09 |
| Execution, Delivery and Process Management (EDPM) | 746 | 0.1 | 0.33 | 9.90 | 5.50 |

From Table 4.1, it can be seen that the EXTF event type has the largest standard deviation, skewness and maximum loss, followed by the CPBP and EDPM event types. Basically, the skewed distribution with a large maximum loss illustrated in Table 4.1 can be caused by either observable or unobservable heterogeneity in each of the event types, the salient empirical regularities of operational risks. Since these events are legitimate operational losses from the bank's business lines with heterogeneous characteristics that have been carefully scrutinized, it only makes sense to include them in the model development and capital estimation.

Table 4.2 displays the five largest events for each event type in the modelling dataset.

| Table 4.2: Five Largest Loss Events | | |
|---|---|---|
| EXTF | CPBP | EDPM |
| 30.93 | 17.09 | 5.50 |
| 0.83 | 8.50 | 3.77 |
| 0.76 | 2.13 | 2.73 |
| 0.66 | 1.07 | 2.65 |
| 0.63 | 0.42 | 1.90 |

For examples, the largest EXTF loss event is caused by Madoff's Ponzi scheme dating back to 2008. The bank's multiple business lines were involved with the event loss amount 37 times of the size of the second event, or 309 times of the size of the average EXTF event. The second largest event is caused by misrepresentation on borrower's applications in broker business, while the third largest event involves data breach that impacts the bank's retail business. Their sizes are 8 times of the average EXTF event. Thus, the EXTF heavy tail is mainly caused by Madoff's Ponzi scheme. For CPBP, the largest loss event is due to anti-money laundering regulatory fines. It also involves the bank's multiple business lines with the loss twice the size of the second largest event in the retail business line, or 171 times of the size of the average CPBP event. The second largest loss event is inherited from a legacy M&A deal from a third party bank, while the third largest loss event is from the retail business line. Finally, the EDPM event type has a regular or light tail when compared to the EXTF and CPBP event types. Its largest event loss amount is 55 times of the size of the average loss event, and its standard deviation and skewness are also smaller than that of the EXTF and CPBP event types.

*4.2. Body-tail threshold*

Following industry practice using the EVT-POT approach with spliced distributions, the body-tail threshold is first identified by finding the best fit Pareto model to the tail event exceedance. It is primarily based on identifying the point with minimum Anderson-Darling (AD) statistic which gives more weight to the tail than Kolmogorov-Smirnov (KS) and Cramer-von Mises (CvM) statistics. Figure 4.1 displays the three test statistics for each of the event types in the dataset.

Figure 4.1 KS, CvM and AD test statistics for the body-tail threshold decision.

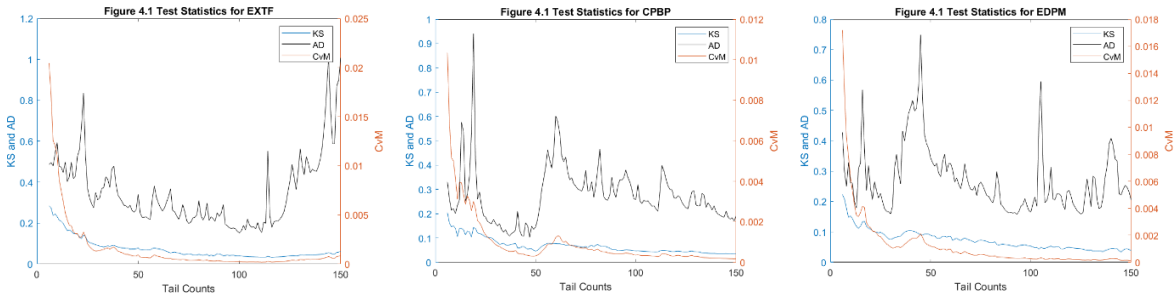

Based on Figure 4.1, the AD statistic suggests that the threshold can be set at 100, 43, and 93 for the EXTF, CPBP and EDPM event types, respectively. From the EXTF, CPBP and EDPM body-tail loss perspective, the tail loss consists of 96%, 99%, and 73% of the total loss, and the body loss consists of 4%, 1%, and 27% of the total loss, respectively. It also indicates that the EDPM event type has a regular or light tail, when compared to the EXTF and CPBP event types. Table 4.3 below illustrates that the KS, CvM and AD tests failed to reject the null hypothesis that the Pareto model provides a good fit to the tail events with a large p-Value for each of the EXTF, CPBP and EDPM event types in the dataset.

| Table 4.3 GPD Model Fit Tests | | KS | CvM | AD |
|---|---|---|---|---|
| EXTF 100 tail events | Statistic | D | W-Sq | A-Sq |
| | | 0.0404 | 0.0204 | 0.1534 |
| | p Value | Pr > D | Pr > W-Sq | Pr > A-Sq |
| | | 0.940 | 0.948 | 0.966 |

|  | Statistic | D | W-Sq | A-Sq |
|---|---|---|---|---|
| CPBP 43 tail events |  | 0.0646 | 0.0135 | 0.1125 |
|  | p-Value | Pr > D | Pr > W-Sq | Pr > A-Sq |
|  |  | 0.888 | 0.996 | 0.998 |
| EDPM 93 tail events | Statistic | D | W-Sq | A-Sq |
|  |  | 0.0535 | 0.0247 | 0.1545 |
|  | p Value | Pr > D | Pr > W-Sq | Pr > A-Sq |
|  |  | 0.708 | 0.900 | 0.976 |

Note that the two parameters Log-logistic (LogLGT) and Lognormal (LGN) models in the Fréchet and Gumbel MDAs are also capable of accommodating heavy tails. Thus to compare their model performance, Table 4.4 below lists log-likelihood values, parameter estimates and quantile estimates at 99.9% for the GPD, LGN and LogLGT models for the tail events.

| Table 4.4 | CPBP | | | EXTF | | | EDPM | | |
|---|---|---|---|---|---|---|---|---|---|
| Models | GPD | LogLGT | LGN | GPD | LogLGT | LGN | GPD | LogLGT | LGN |
| LogL Value | 42.39 | 41.7 | 41.33 | 164.66 | 164.56 | 163 | 11.33 | 11.04 | 11.46 |
| Quantile at 99.9% | 1013.9 | 143.21 | 35.67 | 46.77 | 30.35 | 8.46 | 29.61 | 68.37 | 18.72 |
| Parameter1* | 1.5858 | -3.2692 | -3.1454 | 1.1178 | -3.7057 | -3.6755 | 0.7057 | -1.9741 | -2.0048 |
| Parameter2* | 0.0281 | 1.1921 | 2.1745 | 0.0232 | 1.0306 | 1.8806 | 0.1608 | 0.8975 | 1.5968 |
| t Stat (Parameter1) | 4.0506 | -10.375 | -9.4852 | 5.313 | -20.755 | -19.5446 | 3.9353 | -12.2103 | -12.1077 |
| t Stat (Parameter2) | 2.9106 | 7.8364 | 9.1104 | 4.8931 | 11.9596 | 14.0357 | 5.1603 | 11.5305 | 13.5277 |

* For GPD, Parameter1 is for shape parameter and Parameter2 is for scale parameter. For LGN and LogLGT, Parameter1 is for mean and Parameter2 is for variance. For LogLGT, its shape parameter is inverse of Parameter2, while its scale parameter is exponential of Parameter1.

It can be seen that the GPD model provides a good fit to both of the EXTF and CPBP event types. However, their shape parameter estimates are larger than one, indicating infinite mean and/or variance. As a result, the CPBP quantile estimate at 99.9% in the GPD model is 1,013.9, which is much larger than the quantile estimates from both the LGN model (35.67) and the LogLGT model (143.21), as well as the observed maximum loss of 17.1 in the dataset. Thus, the EXTF and CPBP event types have either heavy or extremely heavy tails. For EDPM, on the other hand, the LGN model outperforms the GPD model, which is likely due to the fact that EDPM has a regular or light tail with the Pareto shape parameter estimate equal to 0.7057, when compared to EXTF and CPBP. Based on Table 4.4, the LGN model in the Gumbel MDA provides the best fit to the EDPM event type. Note that the quantile estimates at 99.9% vary significantly for the parametric GPD, LGN and LogLGT models due to sensitivity of the quantile estimates at 99.9% with respect to the distribution assumptions. Therefore, it can be seen that naïve application of the EVT-POT approach can lead to a counter-intuitive result with infinite mean and/or variance, thus non-admissible capital estimation (Moscadelli, 2004 and Neslova *et* *al.*, 2006). There is a need to search for modeling methodology that can yield an intuitive and stable capital estimate in the presence of heavy tails, which should not be overly sensitive to the selected model distribution and with a certain degree of stability.

## 4.3. SNP models for tail events

In this research, we propose to estimate the operational risk tail loss events from three Basel event types EXTF, CPBP and EDPM using the SNP methodology as defined by equations (1), (2), (3) and probability density function $g(x)$ in equation (5). Based on Section 3, we observed that Pareto, Log-Logistic, and Lognormal distributions in the Fréchet and Gumbel MDAs can be used as kernel distributions in the SNP models to accommodate heavy tails. The distributions in the Weibull MDA may not be the best choice in the presence of heavy tails. After evaluating several SNP models in the Fréchet and Gumbel MDAs with two, three and four additional parameters, we found overall that the SNP models with three additional parameters perform the best for the dataset, especially using the Lognormal kernel distribution. Table 4.5 below presents the goodness of fit, parameter estimates, and 99.9% quantiles for the SNP models with three additional parameters $\theta$s in (3).

| Table 4.5 | | CPBP | | | EXTF | | | EDPM | | |
|---|---|---|---|---|---|---|---|---|---|---|
| Models | | SNPGPD 3p | SNPLGT 3p | SNPLGN 3p | SNPGPD 3p | SNPLGT 3p | SNPLGN 3p | SNPGPD 3p | SNPLGT 3p | SNPLGN 3p |
| LogL Value | | 45.66 | 45.29 | 45.93 | 168.27 | 168.27 | 168.62 | 12.29 | 12.31 | 12.65 |
| Quantile at 99.9% | | 23.20 | 21.00 | 18.90 | 36.10 | 36.38 | 33.23 | 6.67 | 6.94 | 6.08 |
| Parameters | $c$ | 1.5110 | 0.9162 | 1.8787 | 0.9767 | 1.0020 | 1.7630 | 0.6479 | 1.0787 | 1.5357 |
| | $\theta_0$ | 5.8828 | 5.2611 | 5.1856 | 6.3847 | 6.4235 | 6.4332 | 2.5207 | 2.6637 | 2.8344 |
| | $\theta_1$ | -0.0512 | -2.4337 | -2.3907 | -1.6976 | -1.7040 | -1.6965 | -0.5440 | 0.1579 | -1.3110 |
| | $\theta_2$ | -0.3419 | 0.3948 | 0.3725 | 0.1249 | 0.1254 | 0.1226 | 0.1580 | -0.0601 | 1.1686 |
| | $\theta_3$ | 0.0227 | -0.0176 | -0.0161 | -0.0027 | -0.0027 | -0.0026 | 0.0121 | 0.0634 | -0.2315 |
| t-Statistics | $c$ | 3.0861 | 7.7492 | 8.9210 | 4.7965 | 11.9634 | 14.0695 | 1.1789 | 7.0847 | 8.8886 |
| | $\theta_0$ | 5.6268 | 6.8553 | 6.9400 | 9.8443 | 11.5131 | 11.3360 | 9.2939 | 9.1284 | 9.8604 |
| | $\theta_1$ | -0.0180 | -3.2068 | -3.4846 | -3.1340 | -3.1130 | -3.6073 | -0.2600 | 0.0770 | -0.9347 |
| | $\theta_2$ | -0.4770 | 2.6941 | 2.9223 | 2.4001 | 2.3838 | 2.7394 | 0.1603 | -0.0484 | 1.4927 |
| | $\theta_3$ | 0.6196 | -2.5005 | -2.7456 | -2.1923 | -2.1834 | -2.4838 | 0.0734 | 0.2962 | -1.8856 |

By comparing with the model's log-likelihood values between Tables 4.4 and 4.5, it can be seen that the SNP models for CPBP and EXTF can improve the goodness of fit at 10% significance level using the nested LR tests. In particular, the SNPLGN3p models outperform the other SNP models together with significant t-Statistics. However, the SNP model's improvement for the EDPM event type is insignificant in the log-likelihood values. The SNP parameter t-Statistics are also insignificant. This is because the EDPM event type has a light tail with Pareto shape parameter less than one when compared to the EXTF and CPBP event types.

To further compare the model performance, we evaluated the model's Q-Q plots and tail distributions for the CPBP, EXTF and EDPM event types in the following.

Figure 4.3a CPBP model Q-Q plots

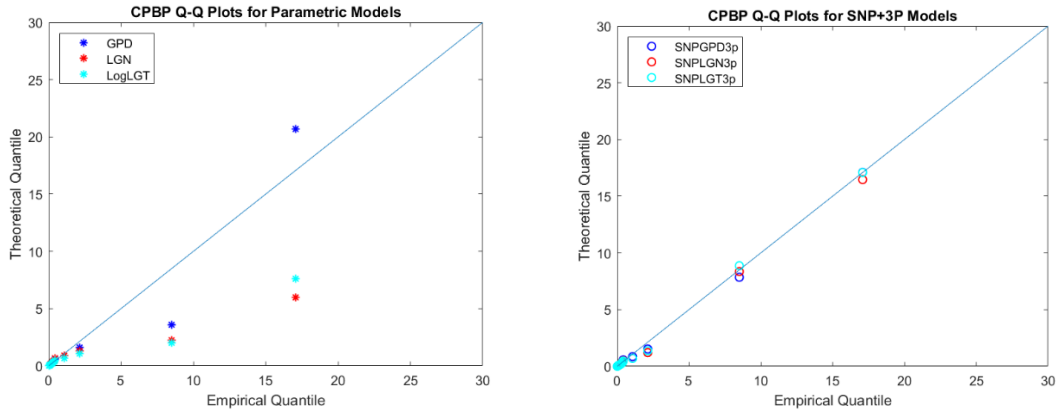

It can be seen in Figure 4.3 that the Q-Q plots for the Pareto model tail events are closer to the 45-degree line than that of the Lognormal and Log-Logistic models. By comparison, the SNP models can further improve the goodness of fit with the Q-Q plots moving further towards the 45-degree line. It appears that all of the SNPLGN3p, SNPGPD3p and SNPLGT models have provided a reasonably good fit to the dataset. The SNP models with four additional parameters are also estimated with performance similar to the SNP models with three additional parameters, thus are omitted here.

Figure 4.3 below indicates that tail behaviors of the three parametric models are different, which may lead to different capital estimates. For example, the CPBP Pareto model has the heaviest tail distribution than that of the Log-Logistic model and the Lognormal model. Based on Table 4.4, the shape parameters are calculated as $c=1.5858$ and $1/\sigma = 0.8389$, respectively, for the Pareto and Log-Logistic models in the Fréchet MDA. The shape parameter is zero for the Lognormal model in the Gumbel MDA, which can be viewed as the limiting case.

Figure 4.3b CPBP model tail distributions

- 23 -

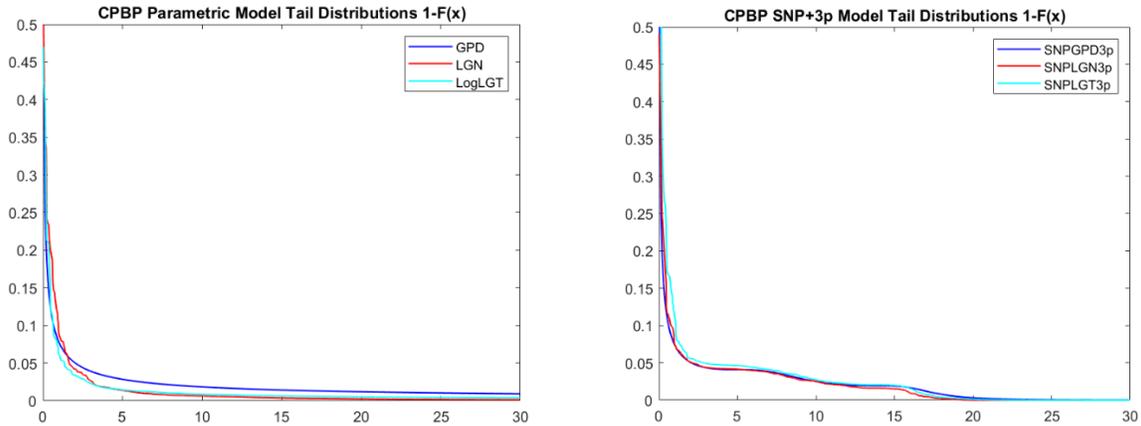

On the other hand, the SNPGDP3p, SNPLGN3p and SNPLGT3p models appear to have a similar tail behavior. Based on *Result* 2 in Section 2, the Fréchet MDA's shape parameters are calculated using Table 4.5 as $\xi = c/m = 0.2157$ and $\xi = 1/(c*m) = 0.1559$ for the SNPGPD3p and SNPLGT3p models, respectively. As a result, the SNP model tails are more stable than the parametric models with $L(x)x^{-1/\xi}$ as $x \to \infty$ asymptotically. This is because the term $x^{-1/\xi}$ or $\xi$ in the SNP models is much smaller than the parametric models, where $L(x)$ is a slow varying function.

For EXTF, the nested LR tests by comparing the log-likelihood values between Tables 4.4 and 4.5 also indicate that the SNP models can improve the goodness of fit over the parametric models at 5% significance level with three degrees of freedom. In particular, the SNPLGN3p model outperforms all of the other models, together with significant t-Statistics for the parameter estimates. Although the KS, CvM, and AD tests in Table 4.3 suggest that the Pareto model can be used to fit the EXTF tail events, the Q-Q plots in Figure 4.4 below indicate that all of the three parametric models clearly under predict the extreme loss event (Madoff Ponzi scheme).

Figure 4.4a EXTF model Q-Q plots.

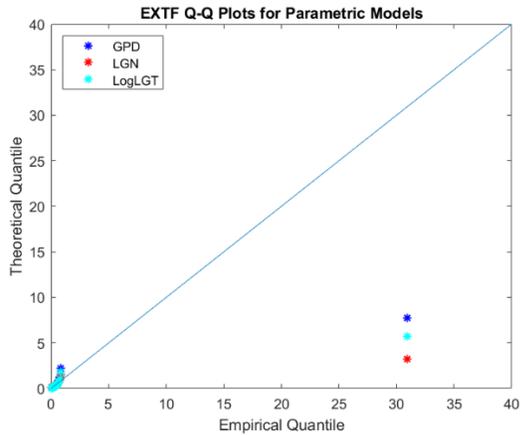
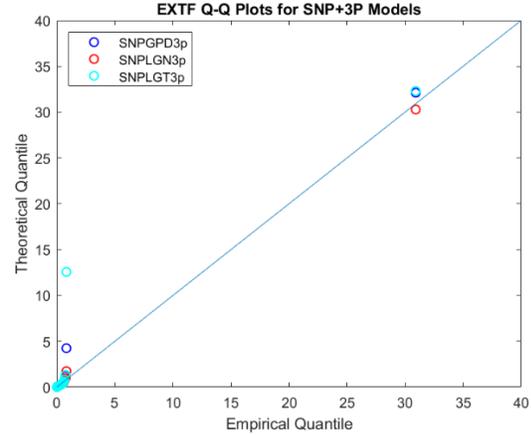

Based on Table 4.4, the EXTF shape parameters for the Pareto and Log-Logistic models are $c=1.1178$ and $1/\sigma = 0.9703$, respectively. Since they are very close to one, the parametric models might not be suitable to the EXTF tail loss events. On the other hand, the Q-Q plots in Figure 4.4 suggest that the SNP models could be a better choice than the parametric models. Similar to CPBP and by using Table 4.5, the shape parameters in the Fréchet MDA based on *Result 2* in Section 2 are calculated as $\xi = c/m = 0.1395$ and $\xi = 1/(c*m) = 0.1426$ for the SNPGPD3p and SNPLGT3p models, respectively.

Figure 4.4b EXTF model tail distributions.

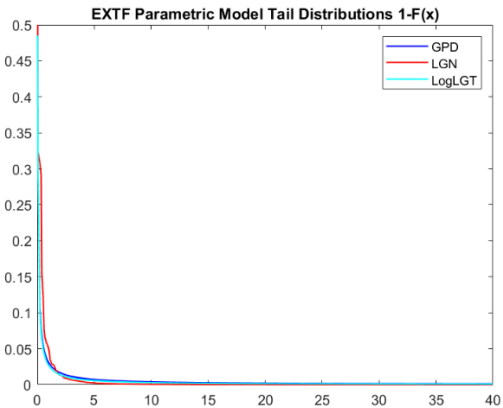
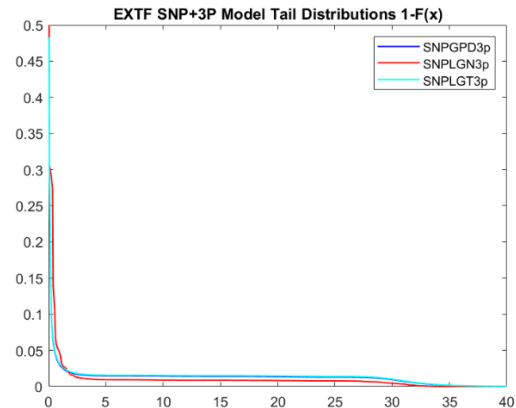

For EDPM, Figure 4.5 indicates that the SNP model's Q-Q plots are along the 45-degree line, which are better than the parametric models. However, the nested LR tests based on Tables 4.3 and 4.4 indicate that the SNP model's improvement is not statistically significant. Furthermore, the SNP parameter t-Statistics are not statistically significant either. Therefore, the Lognormal model appears to be acceptable in this case.

Figure 4.5a EDPM model Q-Q plots.

- 25 -PUBLIC

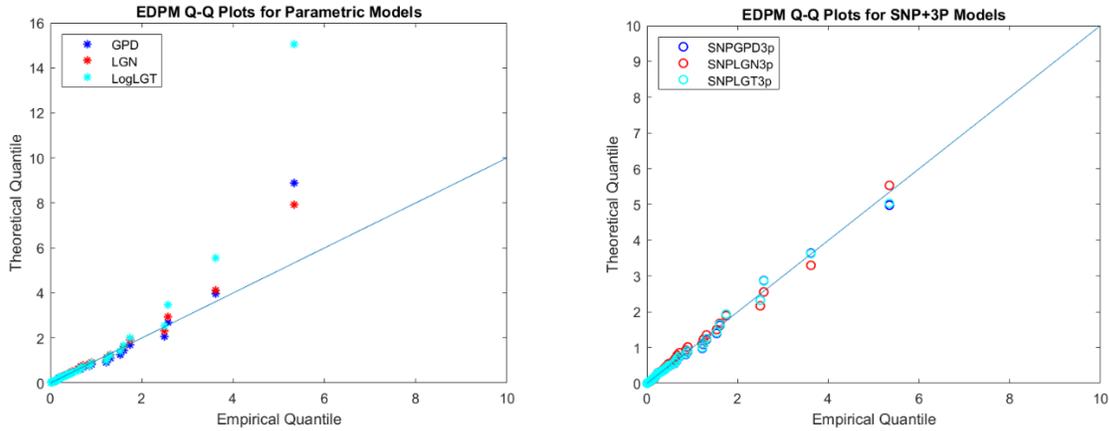

Figure 4.5b EDPM model tail distributions.

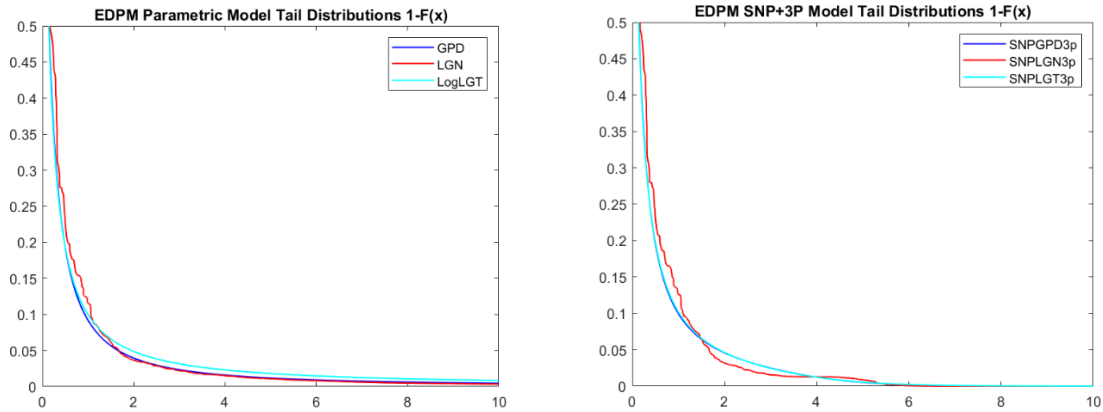

### 4.4. Truncated lognormal distribution for body events

Since there is a minimum reporting threshold of 10,000 USD ($rt$) for bank operational risk losses, the Lognormal distribution for body events will be truncated between the reporting threshold $rt$ and the body-tail threshold $bt$ identified in the previous section. Table 4.6 below presents the model estimates for the body loss events in EXTF, CPBP and EDPM.

| Table 4.6 | EXTF | | CPBP | | EDPM | |
|---|---|---|---|---|---|---|
| Parameter | Estimate | t Statistic | Estimate | t Statistic | Estimate | t Statistic |
| $\mu$ | -5.8952 | -19.97 | -10.062 | -4.46 | -6.4002 | -3.10 |
| $\sigma$ | 0.851 | 4.13 | 2.9345 | 2.59 | 2.498 | 2.77 |

Here ($\mu, \sigma$) are mean and variance of the truncated Lognormal distribution with thresholds ($rt, bt$):

$$\tilde{f}(x|\mu,\sigma) = \frac{f(x|\mu,\sigma)}{F(bt|\mu,\sigma) - F(rt|\mu,\sigma)} \quad (7)$$

where $F$ is Lognormal cumulative distribution, and $f(x|\mu,\sigma)$ is Lognormal density function:

$$f(x|\mu, \sigma) = \frac{1}{x\sigma\sqrt{2\pi}} \exp\left\{-\frac{(\ln(x)-\mu)^2}{2\sigma^2}\right\}$$

The final aggregate loss distribution is estimated by convolution of the spliced distributions for loss severity and Poisson distributions for loss frequency per industry practice.

*4.5. Capital Estimation*

After combining the body and tail events for each of the event types, Table 4.7 compares the annual capital VaR at 99.9% with 10,000 iterations between the three parametric and three SNP models. The observed total loss and largest loss are also presented in the table for the purpose of comparison with the VaR capital estimate. The Poisson parameter estimates for the EXTF, CPBP and EDPM body events are 34.9, 30.4 and 70.6, respectively. The Poisson parameter estimates for the EXTF, CPBP and EDPM tail events are 10.8, 4.6 and 10.1, respectively. Note that the body models use the same truncated Lognormal distribution for the Basel event types in Table 4.6. The log-likelihood value in Table 4.7 is for the tail models.

| Table 4.7 | EXTF | | CPBP | | EDPM | |
|---|---|---|---|---|---|---|
| Total Loss | 42.3 | | 32.4 | | 74.6 | |
| Largest Loss | 30.9 | | 17.1 | | 5.5 | |
| Model Estimates | VaR at 99.9% | Log Likelihood | VaR at 99.9% | Log Likelihood | VaR at 99.9% | Log Likelihood |
| GPD | 983.4 | 164.66 | 10492.5 | 42.39 | 196.5 | 11.33 |
| LogLGT | 506.8 | 164.56 | 830.8 | 41.70 | 741.7 | 11.04 |
| LGN | 36.3 | 163.00 | 87.8 | 41.33 | 71.9 | 11.46 |
| SNPGPD3p | 73.5 | 168.27 | 40.8 | 45.66 | 20.0 | 12.29 |
| SNPLGT3p | 74.6 | 168.27 | 39.7 | 45.29 | 20.3 | 12.31 |
| SNPLGN3p | 66.4 | 168.62 | 36.5 | 45.93 | 20.0 | 12.65 |

From Table 4.7, it can be seen that the SNP models provide an improved fit to the CPBP and EXTF datasets over the parametric models, especially the SNPLGN3p model with the nested LR test exceeding 10% significance level with three degrees of freedom. The SNP model parameter estimates are also highly significant. After comparing the Q-Q plots, the SNPLGN3p is selected as the best model for the capital estimation because its capital estimate is intuitive and sensible. The CPBP SNPLGN3p model capital estimate (36.5) is slightly larger than the observed total loss (32.4), or 2.1 times of the observed largest loss (17.1). The EXTF SNPLGN3p model capital estimate (66.4) is 1.6 times of the observed total loss (42.3), or 2.1 times of the observed largest loss (30.9). On the other hand, the EXTF and CPBP capital estimates from the parametric models are counter-intuitive with the Pareto tail shape parameters larger than one (Table 4.4). Their capital estimates are not only much larger than either the observed total loss or the largest loss in the dataset but also overly sensitive to the parametric distribution assumptions.

For EDPM, since its tail is lighter than that of EXTF and CPBP as discussed in Section 4.1, the parametric models can fit the dataset quite well with the Pareto shape parameter equal to 0.7057. The SNP model's additional parameters are statistically insignificant, and their log-likelihood values are not much better than the parametric models. As a result, the best Lognormal model capital estimate (71.9) should be selected, which is similar to the observed total loss (74.6), or 13 times of the largest loss (5.5).

## 5. Final Comments

In this research, we extend the SNP estimation using Jacobian transformation to estimate the operational risk VaR capital in the EVT-POT approach. The flexible SNP density function consists of a mixture of kernel density functions, and it enriches the family of distributions that can be used to handle heavy tails with shape parameter being a function of the polynomial series order $m$ and the chosen kernel shape parameter $c$. We prove that the SNP distribution has the same MDA as the chosen parametric kernel, and it follows that the SNP model is consistent with the EVT-POT approach. The SNP model is unlike the weighted MLE or other robust estimators in the literature in which the model estimation could be sensitive to the number of assumed outliers and/or tuning parameters. Therefore, the SNP model is robust against gaming since there is no outlier assumption or tuning parameter to manipulate. The extreme loss events are legitimate, which have been scrutinized by bank risk managers and modeling communities. This research suggests that the SNP model may enable banks to continue with AMA or its partial use to obtain an intuitive operational risk capital estimate that is sensitive to risk exposure when the simple non-model based Basic Indicator Approach (BIA) or Standardized Approach (TSA) are not suitable per Basel Committee Banking Supervision OPE10 (2019).

To empirically evaluate the SNP model's advantages, in this research we have estimated the SNP models using both the simulated dataset and actual operational risk loss dataset with both light and heavy tails from a large international bank. The SNP model performance can be improved satisfactorily through suitably increasing the number of model parameters or the truncation point to accommodate the salient empirical regularities of heavy tails in the operational risk. Unlike the popular parametric models, the SNP model quantile estimates at 99.9% are quite insensitive towards the body-tail threshold determination, and the SNP model VaR capital estimates are also intuitive in spite of heavy tails. The SNP model specification is also easy to implement since maximum likelihood estimation can yield the model parameter estimates in just one step. Any conventional software with a suitable optimization routine can be used in the estimation. For the event type with heavy tails, selection of kernel distributions from the Fréchet or Gumbel MDAs will be a good starting point. The kernel distributions from the Weibull MDA may be used for the other event types with light tails. The model's goodness of fit can be improved by gradually increasing the number of parameters in the power series in (3). Since the SNP model nests the chosen parametric model as a special case, the model specification tests can be performed with the conventional test statistics such as the nested LR test and/or Student t-test. Given the vital importance of tail events in the operational risk capital, the Q-Q plots are also shown to be a useful tool in the estimation to ensure that the predicted and the observed are as close as possible.

**Appendix:**

The simulated dataset is generated from a mixture of three distributions, each with 1000 observations. Heavy tails mainly come from Log-Logistic and Pareto distributions with large skewness and standard deviation statistics.

| Table A1 Summary Statistics | Shape | Scale | Sample Size | Minimum | Mean | Maximum | Standard Deviation | Skewness |
|---|---|---|---|---|---|---|---|---|
| Weibull | 5/3 | 1/3 | 1000 | 0 | 12.67 | 978.28 | 55.66 | 11.02 |
| Pareto | 4/3 | 1/4 | 1000 | 0 | 14.57 | 7,856 | 259.13 | 28.15 |
| Log-Logistic | 2/3 | 1/20 | 1000 | 0 | 21.65 | 16,659 | 534.75 | 30.31 |

The following table presents the estimated sixty-three model results with respect to the log-likelihood values, parameter estimates, and t-statistics for each of the twenty-one models based on the three dataset with cutoff thresholds at 50, 30, and 10.

| Table A2 Model Estimates | Threshold Cut at 50 | | | Threshold Cut at 30 | | | Threshold Cut at 10 | | |
|---|---|---|---|---|---|---|---|---|---|
| **Models** | logL | Parm | t-Stats | logL | Parm | t-Stats | logL | Parm | t-Stats |
| **EXP** | -27.88 | 0.54181 | 8.485 | 7.20 | 0.34609 | 10.863 | 197.12 | 0.17136 | 16.062 |
| **GPD** | 44.30 | 1.04267 | 4.614 | 146.47 | 1.17947 | 5.791 | 532.96 | 1.09399 | 8.393 |
| | | 0.07010 | 4.478 | | 0.03269 | 5.119 | | 0.01561 | 7.850 |
| **LogLGT** | 44.69 | -2.64757 | -14.134 | 146.32 | -3.32825 | -19.605 | 532.81 | -4.11285 | -37.108 |
| | | 0.91356 | 10.198 | | 1.06037 | 13.038 | | 1.02727 | 19.198 |
| **LGN** | 43.90 | -2.52763 | -12.932 | 146.12 | -3.29295 | -18.862 | 530.51 | -4.09057 | -35.440 |
| | | 1.65847 | 11.874 | | 1.89642 | 15.264 | | 1.85396 | 22.650 |
| **WBL** | 29.55 | 0.19271 | 4.105 | 128.87 | 0.09833 | 4.911 | 491.23 | 0.04319 | 7.294 |
| | | 0.51508 | 12.688 | | 0.48042 | 16.075 | | 0.48234 | 24.322 |
| **GB2** | 47.83 | 0.75564 | 1.035 | 147.17 | 0.40545 | 1.249 | 533.15 | 0.74487 | 1.838 |
| | | 0.00269 | 0.157 | | 0.00376 | 0.266 | | 0.01092 | 1.505 |
| | | 8.83096 | 0.245 | | 6.47841 | 0.517 | | 1.67811 | 1.060 |
| | | 1.08764 | 0.775 | | 2.84780 | 0.795 | | 1.33843 | 1.294 |
| **SNPEXP2p** | 29.91 | 2.22311 | 16.957 | 101.98 | 2.69790 | 21.715 | 393.74 | 0.56549 | 22.569 |
| | | -0.64677 | -10.237 | | -0.78543 | -12.354 | | 5.05849 | 17.115 |
| | | 0.03584 | 7.882 | | 0.04334 | 9.355 | | -1.52032 | -8.924 |
| **SNPGPD2p** | 46.61 | 0.71898 | 3.553 | 147.90 | 0.95394 | 5.011 | 534.53 | 0.97602 | 7.766 |
| | | 3.57564 | 9.169 | | 5.28019 | 10.406 | | 7.81090 | 15.837 |
| | | -1.11028 | -5.958 | | -1.63781 | -6.250 | | -2.41218 | -7.049 |
| | | 0.06499 | 4.541 | | 0.09612 | 4.610 | | 0.14125 | 4.916 |
| **SNPLGT2p** | 47.46 | 1.19736 | 10.314 | 147.88 | 0.99211 | 13.108 | 534.51 | 0.99946 | 19.246 |
| | | -3.84639 | -11.537 | | -5.34798 | -12.310 | | -7.85971 | -18.455 |
| | | 1.20065 | 6.430 | | 1.65671 | 6.520 | | 2.42502 | 7.171 |
| | | -0.07055 | -4.724 | | -0.09725 | -4.705 | | -0.14200 | -4.947 |

- 30

| Model | | | | | | | | |
|---|---|---|---|---|---|---|---|---|
| SNPLGN2p | 46.85 | 1.51170 | 11.573 | 148.34 | 1.80557 | 14.976 | 533.70 | 1.79876 | 22.439 |
| | | -3.66224 | -11.177 | | -5.29741 | -11.997 | | -7.81928 | -17.839 |
| | | 0.65445 | 3.628 | | 0.90770 | 3.098 | | 1.40316 | 3.794 |
| | | -0.03909 | -2.499 | | -0.05595 | -2.198 | | -0.08451 | -2.500 |
| SNPWBL2p | 41.67 | 0.68822 | 11.669 | 142.55 | 0.59428 | 14.836 | 514.41 | 0.56549 | 22.569 |
| | | 2.59220 | 10.989 | | 3.48249 | 12.170 | | 5.05849 | 17.115 |
| | | -0.77224 | -7.205 | | -1.04251 | -7.509 | | -1.52032 | -8.924 |
| | | 0.04371 | 5.681 | | 0.05911 | 5.808 | | 0.08589 | 6.663 |
| SNPGPD3p | 47.33 | 0.91868 | 2.840 | 149.36 | 1.21393 | 3.590 | 535.88 | 1.08188 | 6.608 |
| | | 3.69618 | 8.639 | | 5.54449 | 9.340 | | 7.97482 | 15.102 |
| | | -0.25240 | -0.194 | | 1.14527 | 0.289 | | 0.42462 | 0.131 |
| | | -0.15543 | -0.465 | | -0.62388 | -0.602 | | -0.57909 | -0.700 |
| | | 0.01130 | 0.650 | | 0.03723 | 0.685 | | 0.03687 | 0.857 |
| SNPLGT3p | 48.10 | 1.15204 | 9.366 | 149.27 | 0.95115 | 11.753 | 535.81 | 0.97953 | 18.165 |
| | | -3.80324 | -11.029 | | -5.28134 | -11.739 | | -7.82167 | -18.070 |
| | | 0.56752 | 0.727 | | -0.10667 | -0.061 | | 0.14362 | 0.064 |
| | | 0.08407 | 0.447 | | 0.34095 | 0.774 | | 0.42599 | 0.762 |
| | | -0.00779 | -0.810 | | -0.02228 | -0.975 | | -0.02885 | -0.996 |
| SNPLGN3p | 49.13 | 1.42773 | 11.257 | 150.32 | 1.78864 | 14.077 | 535.93 | 1.78158 | 21.796 |
| | | -3.76535 | -11.739 | | -5.31855 | -11.957 | | -7.85081 | -17.929 |
| | | 1.49475 | 3.717 | | 0.96686 | 1.117 | | 1.86715 | 1.928 |
| | | -0.22811 | -2.905 | | 0.06451 | 0.329 | | -0.01064 | -0.049 |
| | | 0.01010 | 2.721 | | -0.00789 | -0.811 | | -0.00635 | -0.597 |
| SNPWBL3p | 42.16 | 0.71900 | 10.821 | 143.19 | 0.61411 | 13.992 | 516.71 | 0.58475 | 21.686 |
| | | 2.66136 | 11.309 | | 3.55021 | 12.476 | | 5.13504 | 17.653 |
| | | -1.13749 | -5.426 | | -1.53559 | -5.751 | | -2.36733 | -7.888 |
| | | 0.16208 | 4.260 | | 0.22036 | 4.426 | | 0.33985 | 5.867 |
| | | -0.00669 | -3.922 | | -0.00913 | -4.026 | | -0.01389 | -5.173 |
| SNPEXP3p | 33.62 | 2.38843 | 16.211 | 110.84 | 2.91530 | 21.152 | 425.54 | 3.98870 | 31.900 |
| | | -0.97406 | -7.530 | | -1.27059 | -10.444 | | -1.98508 | -19.661 |
| | | 0.09943 | 4.416 | | 0.13902 | 6.535 | | 0.25497 | 13.402 |
| | | -0.00283 | -2.915 | | -0.00428 | -4.673 | | -0.00941 | -10.496 |
| SNPGPD4p | 47.34 | 0.89721 | 2.652 | 149.36 | 1.20853 | 3.471 | 535.88 | 1.07623 | 6.393 |
| | | 3.69050 | 8.614 | | 5.54164 | 9.310 | | 7.96860 | 15.053 |
| | | -0.46490 | -0.291 | | 0.99751 | 0.220 | | 0.10600 | 0.027 |
| | | -0.07116 | -0.137 | | -0.55917 | -0.390 | | -0.43825 | -0.330 |
| | | 0.00186 | 0.037 | | 0.02922 | 0.213 | | 0.01937 | 0.137 |
| | | 0.00031 | 0.185 | | 0.00029 | 0.062 | | 0.00063 | 0.125 |
| SNPLGT4p | 48.18 | 1.16411 | 9.170 | 149.28 | 0.95357 | 11.559 | 535.83 | 0.98135 | 17.954 |
| | | -3.82229 | -11.066 | | -5.28864 | -11.710 | | -7.82763 | -18.066 |
| | | 0.89223 | 0.859 | | 0.10767 | 0.049 | | 0.52158 | 0.184 |
| | | -0.05445 | -0.148 | | 0.24282 | 0.318 | | 0.25082 | 0.249 |
| | | 0.00902 | 0.213 | | -0.00966 | -0.112 | | -0.00622 | -0.053 |
| | | -0.00060 | -0.384 | | -0.00047 | -0.147 | | -0.00084 | -0.192 |
| **SNPLGN4p** | 49.61 | 1.37962 | 10.990 | 150.35 | 1.78114 | 13.649 | 536.06 | 1.77268 | 21.422 |
| | | -3.83148 | -12.066 | | -5.33296 | -11.894 | | -7.87248 | -17.954 |
| | | 2.16361 | 3.954 | | 1.18869 | 0.925 | | 2.46243 | 1.729 |
| | | -0.54313 | -2.978 | | -0.02404 | -0.055 | | -0.25331 | -0.510 |
| | | 0.05287 | 2.579 | | 0.00222 | 0.048 | | 0.02162 | 0.395 |
| | | -0.00169 | -2.374 | | -0.00034 | -0.220 | | -0.00095 | -0.504 |
| **SNPWBL4p** | 42.82 | 0.75084 | 10.291 | 143.87 | 0.63223 | 13.453 | 518.58 | 0.59991 | 21.174 |
| | | 2.72003 | 11.652 | | 3.60299 | 12.751 | | 5.19074 | 18.076 |
| | | -1.43220 | -4.763 | | -1.90453 | -5.020 | | -3.01349 | -7.522 |
| | | 0.27127 | 2.853 | | 0.36275 | 2.927 | | 0.60563 | 4.323 |
| | | -0.01905 | -1.875 | | -0.02554 | -1.900 | | -0.04531 | -2.916 |
| | | 0.00042 | 1.243 | | 0.00056 | 1.251 | | 0.00109 | 2.086 |
| **SNPEXP4p** | 35.91 | 2.48865 | 15.705 | 115.98 | 3.04936 | 20.815 | 432.30 | 4.10336 | 31.586 |
| | | -1.21316 | -6.488 | | -1.64062 | -10.247 | | -2.38844 | -18.710 |
| | | 0.16955 | 2.906 | | 0.26448 | 5.123 | | 0.42553 | 9.816 |
| | | -0.00862 | -1.370 | | -0.01660 | -2.893 | | -0.02995 | -5.935 |
| | | 0.00012 | 0.585 | | 0.00035 | 1.774 | | 0.00072 | 4.084 |

# References

Abdymomunov A, Curti F (2019) Quantifying and stress testing operational risk with peer banks' data. Journal of Financial Services Research. Published online, 31 July 2019. https://doi.org/10.1007/s10693-019-00320-w.

Alves I, Neves C (2017) Extreme value theory: an introductory overview. In Longin F (eds.), extreme events in finance: a handbook of extreme value theory and its applications. Chapter Four, Pages 53-95, John Wiley & Sons, Inc., ISBN9781118650196.

Basel Committee on Banking Supervision (2006) International convergence of capital measurement and capital standards, https://www.bis.org/publ/bcbs128.htm.

Basel Committee on Banking Supervision (2019) OPE calculation of RWA for operational risk and OPE10 definitions and application, https://www.bis.org/basel_framework/chapter/OPE/10.htm.

Chen H, Randall A (1997) Semi-nonparametric estimation of binary choice models for natural resource valuation, Journal of Econometrics, Vol. 76, Pages 323-240. DOI: https://doi.org/10.1016/0304-4076(95)01794-1.

Chen X (2007) Large sample sieve estimation of semi-nonparametric models. In Heckman J and Leamer E (Eds.), Handbook of Econometrics, Chapter 76, Vol. 6B, Pages 5549-5632, Elsevier B.V. Press. DOI: https://doi.org/10.1016/S1573-4412(07)06076-X.

Choi E, Hall P, Presnell B (2000) Rendering parametric procedures more robust by empirically tilting the model, Biometrika, Vol. 87, No 2, Pages 453-465. DOI: https://doi.org/10.1093/biomet/87.2.453.

Colombo A, Lazzarini A, Mongelluzzo S (2015) A weighted likelihood estimator for operational risk data: improving the accuracy of capital estimates by robustifying maximum likelihood estimates, Journal of Operational Risk, Vol. 10 No. 3. Pages 47-108. DOI: 10.21314/JOP.2015.164.

Cope EW, Mignola G, Antonini G, Ugoccioni R (2009) Challenges in measuring operational risk from loss data. Journal of Operational Risk 4:3-27. DOI: 10.21314/JOP.2009.069.

Dutta K, Perry J (2007) A tale of tails: an empirical analysis of loss distribution models for estimating operational risk capital. Federal Reserve Bank of Boston no. 06-13. DOI: http://dx.doi.org/10.2139/ssrn.918880.

Embrechts P, Klüppelberg C, Mikosch T (1997) Modelling extremal events. Springer-Verlag Press. ISBN 3-540-60931-8.

Franzetti C (2011) Operational risk modeling and management. Chapman & Hall/CRC Finance Series. ISBN 9781138116511 - CAT# K35501.

Gallant A R, Nychka, D W (1987) Semi-nonparametric maximum likelihood estimation. Econometrica Vol. 55, pp 363-390. DOI: 10.2307/1913241.

Horbenko N, Ruckdeschel P, Bae T (2011) Robust estimation of operational risk, Cornell University Library. DOI: https://arxiv.org/abs/1012.0249v3.

Moscadelli M (2004) The modelling of operational risk: experience with the analysis of the data collected by the Basel Committee. Banca D'Italia discussion paper. DOI: http://dx.doi.org/10.2139/ssrn.557214.

Neslova J, Embrechts P, Chavez-Demoulin V (2006) Infinite-mean models and the LDA for operational risk, Journal of Operational Risk, 1(1): Pages 3-25. DOI:10.21314/JOP.2006.001.

Ñíguez T M, and Perote J (2011) A new proposal for computing portfolio value-at-risk for semi-nonparametric distributions.